\documentclass[showpacs,twocolumn,pra,superscriptaddress,notitlepage,showkeys]{revtex4-1}
\usepackage{qcircuit}
\usepackage[dvips]{graphicx}
\usepackage{amsmath,amssymb,amsthm,mathrsfs,amsfonts,dsfont}
\usepackage{subfigure, epsfig}

\usepackage{braket}
\usepackage{bm}
\usepackage{enumerate}
\usepackage{color}

\newcommand{\M}{{M}}

\usepackage{hyperref}
\hypersetup{colorlinks=true, linkcolor=blue, citecolor=black, urlcolor=black}

\begin{document}
\title{Variational algorithms for linear algebra}
\begin{abstract}
Quantum algorithms have been developed for efficiently solving linear algebra tasks. However, they generally require deep circuits and hence universal fault-tolerant quantum computers. In this work, we propose variational algorithms for linear algebra tasks that are compatible with noisy intermediate-scale quantum  devices. We show that the solutions of linear systems of equations and matrix-vector multiplications can be translated as the ground states of the constructed Hamiltonians.
Based on the variational quantum algorithms,  we introduce Hamiltonian morphing together with an adaptive ans\"atz for efficiently finding the ground state, and show the solution verification. 
Our algorithms are especially suitable for linear algebra problems with sparse matrices, and have wide applications in machine learning and optimisation problems. The algorithm for matrix multiplications can be also used for Hamiltonian simulation and open system simulation. 
We evaluate the cost and effectiveness of our algorithm through numerical simulations for solving linear systems of equations. We implement the algorithm on the IBM quantum cloud device with a high solution fidelity of 99.95\%.

\end{abstract}


\date{\today}

\author{Xiaosi Xu}
\affiliation{Center on Frontiers of Computing Studies, Department of Computer Science, Peking University, Beijing 100871, China}
\affiliation{Department of Materials, University of Oxford, Parks Road, Oxford OX1 3PH, United Kingdom}

\author{Jinzhao Sun}
\affiliation{Clarendon Laboratory, University of Oxford, Parks Road, Oxford OX1 3PU, United Kingdom}

\author{Suguru Endo}
\affiliation{Department of Materials, University of Oxford, Parks Road, Oxford OX1 3PH, United Kingdom}
\author{Ying Li}
\affiliation{Graduate School of China Academy of Engineering Physics, Beijing 100193, China}

\author{Simon C.~Benjamin}
\affiliation{Department of Materials, University of Oxford, Parks Road, Oxford OX1 3PH, United Kingdom}

\author{Xiao Yuan}
\email{xiaoyuan@pku.edu.cn}
\affiliation{Center on Frontiers of Computing Studies, Department of Computer Science, Peking University, Beijing 100871, China}
\affiliation{Department of Materials, University of Oxford, Parks Road, Oxford OX1 3PH, United Kingdom}

\maketitle

\section{Introduction}
Quantum computing has wide applications in various linear algebra tasks. A prominent example is the quantum algorithm for solving linear systems of equations by Harrow, Hassidim, and Lloyd (HHL) in Ref.~\cite{PhysRevLett.103.150502}. Given an $N$ by $N$ sparse matrix $\M$ and a state vector $\ket{v_0}$, the HHL quantum algorithm can prepare a state that is proportional to $
\ket{v_{M^{-1}}} = {\M}^{-1} \ket{v_0}$, 
with a complexity polynomial in $\log_2N$ and condition number $\kappa$, which is the ratio of the largest eigenvalue to the smallest eigenvalue of a given matrix. As classical algorithms generally have polynomial complexity in $N$, the HHL algorithm suggests exponential speed-ups of quantum computers. Recent developments of quantum algorithms for linear systems of equations  can be found in Refs.~\cite{doi:10.1137/16M1087072, ambainis2012variable, PhysRevLett.110.250504, PhysRevLett.120.050502, chakraborty2018power, gilyen2018quantum, PhysRevLett.122.060504}.
Another common linear algebra task is matrix-vector multiplication by applying a sparse matrix $\M$ to a vector $\ket{v_0}$ as
$
\ket{v_M} = {\M} \ket{v_0}$.
The quantum algorithms for linear equations can be similarly applied for matrix-vector multiplications. Furthermore, they can be applied for quantum optimisation and machine learning~\cite{PhysRevLett.113.130503,lloyd2014quantum,biamonte2017quantum}.

The conventional quantum algorithms generally require a long-depth circuit for a fault tolerant quantum computer. This is challenging for current technology.
Recently, there has been great interest in quantum computing in the noisy intermediate-scale quantum (NISQ) regime, for finding energy spectra of a many-body Hamiltonian~\cite{peruzzo2014variational,wang2015quantum,PRXH2,PhysRevA.95.020501,VQETheoryNJP,PhysRevLett.118.100503,PhysRevX.8.011021,Santagatieaap9646,kandala2017hardware,kandala2018extending,PhysRevX.8.031022,kokail2018self,yuan2020quantum,sun2021perturbative}, simulating real and imaginary time dynamics of many-body systems~\cite{Li2017,mcardle2018variational,endo2020variational,Fujirealtime19,chen2019demonstration}, applications with machine learning~\cite{romero2017quantum,farhi2018classification,perdomo2018opportunities,benedetti2018generative,havlivcek2019supervised,lamata2018quantum,khoshaman2018quantum,lloyd2018quantum}, circuit learning~\cite{mitarai2018quantum,cincio2018learning,jones2018quantum,biamonte2021universal,sharma2019noise}, and others~\cite{2019arXiv190709032L,lee2019hybrid}. 
These algorithms are generally hybrid in a sense that they only solve the core problem with a shallow quantum circuit and leave the higher level calculation to be performed with a classical computer. Furthermore, even without error correction, noise in the shallow circuit implementation can be suppressed via error mitigation~\cite{subspace1,Li2017,PhysRevLett.119.180509, endo2017practical, subspace2,recoveringnoisefree,PhysRevA.99.012334, samerrormitigation,bonet2018low,sun2020mitigating2}, indicating the feasibility of quantum computing with NISQ hardware. 

In this work, we propose variational algorithms for linear algebra problems, including linear systems of equations and matrix-vector multiplications, implemented with NISQ hardware. The main idea is to construct a many-body Hamiltonian so that its ground state corresponds to the solution of the linear algebra problem. Then we apply the recently proposed variational methods, such as variational quantum eigensolver (VQE)~\cite{peruzzo2014variational,cerezo2020variational}, imaginary time evolution (ITE)~\cite{mcardle2018variational,endo2020variational} and adaptive variational methods~\cite{grimsley2019adaptive,zhang2020low}, to find the ground state and hence the solution.
We also introduce Hamiltonian morphing with an adaptive ans\"atz to address the barren plateau problem~\cite{mcclean2018barren}.
Different from the conventional scenario where the ground state energy is generally unknown, the ground state energy of the linear algebra Hamiltonians is designed to be zero in our algorithms, and this further enables us to verify the correctness of the solution.
Meanwhile, we show that the variational matrix-vector multiplication algorithm can be applied for Hamiltonian simulation as an alternative to the one proposed by Li and Benjamin~\cite{Li2017}. 

The soundness of our proposed algorithm is evaluated via numerical simulations for solving linear systems of equations with random matrices that have sizes up to 64*64. We discover a polynomial scaling of the circuit depth and computation time with increasing matrix size and condition number. Finally we present an experiment implementing our algorithm on the IBM Q device. Considering a 2-dimensional linear system, we show a final state with 99.95\% fidelity with respect to the target state. This simple example is encouraging for the prospects of our algorithm as an option for solving linear algebra problems when the quantum devices with larger size and better statistics become available.

\section{Variational algorithms for linear algebra}

We first introduce our variational  algorithms for  matrix-vector multiplication. Given a sparse $N$ by $N$ matrix $\M$ and an initial state vector $\ket{v_0}$, our task is to calculate the normalised state 
\begin{align} 
\label{Eq:MatrixMulti}
\ket{v_{\M}} = \frac{{\M} \ket{v_0}}{\|{\M} \ket{v_0}\|},
\end{align}
with $\|{\M} \ket{v_0}\| = \bra{v_0}{\M}^\dag {\M} \ket{v_0}$ and $M\ket{v_0}\ne 0$. Here, we consider the case that $\M$ can be a general (non-Hermitian) matrix. We find that $\ket{v_{\M}}$ is the ground state of the Hamiltonian
\begin{align}
H_{{\M}}=I-\frac{{\M}\ket{v_0}\bra{v_0}{\M^\dag}}{\|{\M} \ket{v_0}\|^2},
\end{align}
with energy $0$.
Using universal quantum computers, one may apply the conventional techniques~\cite{Abrams97,aspuru2005simulated}, such as adiabatic state preparation and phase estimation to find the ground state of $H_{{\M}}$.  With NISQ devices, we can instead consider  the variational method with a
parameterised state. The main idea is to first consider a state or ans\"atz, $\ket{\phi(\vec{\theta})}=U(\vec{\theta})\ket{0}$, with $U(\vec{\theta}) = U_L(\theta_L)\dots U_2(\theta_2)U_1(\theta_1)$,  $\vec{\theta}=(\theta_1,\theta_2,...,\theta_L)$. 
Suppose the ground state of $H_{{\M}}$ can be represented by the ans\"atz with certain parameters, the ground state finding problem is then converted to the minimisation problem,
\begin{align}\label{overlap}
\vec{\theta}_{\min}=\arg \min_{\vec{\theta}}\bra{\phi(\vec{\theta})} H_{{\M}} \ket{\phi(\vec{\theta})}.
\end{align}
with the solution given by $\ket{v_{\M}} = \ket{\phi(\vec{\theta}_{\min})}$. We show how to measure the expectation value $\bra{\phi(\vec{\theta})} H_{{\M}} \ket{\phi(\vec{\theta})}$ in the next section and note that the solution given by the variational algorithms satisfies $\braket{\phi(\theta)|H_M|\phi(\theta)}\geq  0$ (See Appendix~\ref{appendix:proof}).
The minimisation can be accomplished with various methods, such as gradient based classical optimisation VQE~\cite{peruzzo2014variational}, global search algorithms~\cite{kokail2018self}, adaptive variational algorithms~\cite{grimsley2019adaptive,zhang2020low}, ITE~\cite{mcardle2018variational,endo2020variational}, etc.
In the main text, we focus on the VQE method and leave the discussion of other methods in Appendix~\ref{appendix:ITE}. With VQE via gradient descent~\cite{peruzzo2014variational}, we start with a guess of the solution $\vec{\theta}_0$, and update the parameters along the negative gradient of the energy $E_{\M}(\vec\theta) = \bra{\phi(\vec{\theta})}H_{{\M}}\ket{\phi(\vec{\theta})}$, 
\begin{equation}
	\vec{\theta}_{i+1} = \vec{\theta}_{i} - a \nabla E_{\M}(\vec\theta_i), \,\forall i  = 0, 1, \dots, T
\end{equation}
where $a$ is the time step, and $T$ is the total number of steps. We show how to measure the gradient $\nabla E_{M}(\vec\theta_i)$ in the next section.

Next, we consider the problem of solving linear equations. Given a sparse $N$ by $N$ matrix $\M$ and an initial state vector $\ket{v_0}$, we want to calculate
\begin{align} 
\label{Eq:MatrixLinear}
\ket{v_{\M^{-1}}} = \frac{\M^{-1} \ket{v_0}}{\|{\M}^{-1} \ket{v_0} \|},
\end{align}
where the state is normalised by $\|{\M}^{-1} \ket{v_0} \| = \sqrt{\bra{v_0}({M}^{-1})^{\dagger} {M}^{-1} \ket{v_0}}$, ${\M}^{-1} \ket{v_0}\ne 0$. When the matrix $\M$ is not Hermitian, it can always be converted into an equivalent problem with a Hermitian matrix by adding one ancillary qubit~\cite{PhysRevLett.103.150502}. We therefore focus on the case where $\M$ is Hermitian and invertible. 
The solution $\ket{v_{\M^{-1}}}$ is the ground state of the Hamiltonian
\begin{align}
H_{{\M}^{-1}}={\M}^\dag (I-\ket{v_0}\bra{v_0}){\M}
\end{align} 
with energy $0$ (See Appendix~\ref{appendix:proof}). Again, one can apply adiabatic algorithms to find the ground state with a universal quantum computer~\cite{PhysRevLett.122.060504}.
Here, we focus on the variational algorithm and consider the following optimisation problem
\begin{align}\label{overlap}
\vec{\theta}_{\min}=\arg \min_{\vec{\theta}}\bra{\phi(\vec{\theta})} H_{{\M}^{-1}} \ket{\phi(\vec{\theta})},
\end{align}
with $\ket{v_{\M^{-1}}} = \ket{\phi(\vec{\theta}_{\min})}$. With VQE, we start with a guess $\vec{\theta}_0$ and update the parameters by
\begin{equation}
	\vec{\theta}_{i+1} = \vec{\theta}_{i} - a \nabla E_{\M^{-1}}(\vec\theta_i), \,\forall i  = 0, 1, \dots, T
\end{equation}
with energy $E_{\M^{-1}}(\vec\theta) = \bra{\phi(\vec{\theta})}H_{{\M}^{-1}}\ket{\phi(\vec{\theta})}$, time step $a$, and total number of steps $T$. 

As the energy $E_{\M}$ or $E_{\M^{-1}}$  monotonically decreases with sufficiently small $a$, the solution via VQE always at least corresponds to a local minimum.  Starting from several random initial positions, one may find the true solution, i.e., the global minimum, whereas there is no guarantee that the true minimum can be reached, especially with a short circuit ans\"atz.  
In conventional VQE, it is generally hard to verify whether the true ground state is achieved, owing to unknown ground state or ground state energy. While for the tasks considered in this work, we can verify the correctness of the final state found, as discussed in the section~\ref{sec:verify}. 

\section{Implementation with quantum circuits}
In this section, we discuss the implementation of our algorithms with quantum circuits. To realise the VQE optimisation, we need to obtain $\nabla E_{\M^{-1}}$ or $\nabla E_{\M}$, which can be estimated either with the finite difference formulae or quantum gradient finding methods recently considered in Refs.~\cite{Li2017, dallaire2018low,romero2018strategies}. 
Here, we only focus on $\nabla E_{\M }$ in the main text and leave the discussion on the implementation of $\nabla E_{\M^{-1}}$ in Appendix~\ref{appendix:circuit}. 

With the finite difference formulae, each component ${\partial E_{\M}}/{\partial \theta_i}$ of the gradient $\nabla E_{\M}=(\partial E_{\M}/\partial \theta_i)$ around $\vec\theta$ can be calculated as
\begin{equation}
	\frac{\partial E_{\M}}{\partial \theta_i} \approx \frac{E_{\M}(\vec{\theta}+\delta \theta_i) - E_{\M}(\vec{\theta}-\delta \theta_i)}{2\delta \theta_i},
\end{equation}
with $\delta\theta_i \ll 1$.
Note that the energy is given by $E_{\M}(\vec\theta)=1-|\bra{\phi(\vec{\theta})}{\M}\ket{v_0}|^2$ with an arbitrary angle $\vec\theta$ (Here we take normalised $M\ket{v_0}$ such as $\|M\ket{v_0}\|=1$). 
Suppose $\M$ can be decomposed as ${\M}=\sum_j \lambda_j \sigma_j$ with unitary operators $\sigma_i$, we can then obtain $E_{\M}(\vec\theta)$  as
\begin{align}
E_{\M}(\vec\theta) = 1 -  \big|\sum_i \lambda_i \bra{\phi(\vec{\theta})}\sigma_i \ket{v_0} \big|^2.
\label{overlap2}
\end{align}
For each term, we denote $\bra{\phi(\vec{\theta})}\sigma_i \ket{v_0}=\braket{0|U|0}$ with $U=U(\vec\theta)^\dag\sigma_iU_{v_0}$ and $\ket{v_0} = U_{v_0}\ket{0}$. The real and imaginary part of $\braket{0|U|0}$ can be evaluated with the Hadamard test or swap test quantum circuits. 
An alternative method is to construct a linear combination of operators to represent $M\ket{v_0}\bra{v_0}M^{\dagger}$ and then we can obtain $E_M$ by measuring the corresponding operators~\cite{biamonte2021universal,wu2021overlapped}. 
 We can evaluate
$E_{\M^{-1}}(\vec\theta)$ in a similar way.

We next discuss the resource estimation for measuring general matrices.
Suppose $\M$ consists of a polynomial number (with respect to the number of qubits) of tensor products of local operators $\sigma_j$, we can then measure each term and efficiently calculate $E_{\M}(\vec\theta)$.
Meanwhile, even if ${\M}=\sum_j \lambda_j \sigma_j$ has exponential number of terms with respect to the number of qubits, the algorithm is still efficient if we can efficiently sample the probability distribution $|\lambda_i|/\sum_i|\lambda_i|$ with a polynomial dependence of $\sum_i|\lambda_i|$ on the number of qubits. This includes cases where the coefficients have analytical expressions, which enables efficient sampling, or the matrix is an oracle-based sparse matrix.
We note that the sample complexity is related to $C = \sum_j {|\lambda_j|}$. According to Hoeffding's inequality, we need $\mathcal O(\log(\delta^{-1})C^2/\varepsilon^2)$ samples to measure $\bra{\phi}M \ket{\psi}$ to an accuracy $\varepsilon\in(0,1)$ with failure probability $\delta$. Specifically, $C$ increases polynomially with respect to the system size.
when $M$ is a sparse matrix with only a polynomial number of non-zeros terms of the matrix elements $
    \{({\bf x}^{(i)}, {\bf y}^{(i)}, M_{{\bf x}^{(i)},{\bf y}^{(i)}})\}
$
where the $(i)$ represents the index of non-zero terms and $M_{{\bf x}^{(i)},{\bf y}^{(i)}}$ represents the matrix element at $({\bf x}^{(i)},{\bf y}^{(i)})$. Suppose $M$ is an $N\times N$ matrix with $n = \lceil\log_2 N\rceil$ and a binary representation of ${\bf x}^{(i)}$ and ${\bf y}^{(i)}$. We can represent $M$ as
$
    M = \sum_{(i)} M_{{\bf x}^{(i)},{\bf y}^{(i)}} \ket{{\bf x}^{(i)}}\bra{{\bf y}^{(i)}}$, which can be expanded as a linear combination of unitary operators. Therefore we can use the quantum circuits in Appendix to calculate  $E_{\M}(\vec\theta)$.
We can also measure the matrices by sampling according to  $q(i) =  M_{{\bf x}^{(i)},{\bf y}^{(i)}} / \sum_{(i)}  M_{{\bf x}^{(i)},{\bf y}^{(i)}} $, which leads to a polynomial resource cost $C = \sum_j {|\lambda_j|} = \sum_{(i)}  M_{{\bf x}^{(i)},{\bf y}^{(i)}} $ for measurements. The cost is small when our assumption holds. 
We refer to Appendix~\ref{appendix:sparse} for more details.

With the quantum gradient finding method, we first express the gradient as
\begin{equation}
    \frac{\partial E_{\M}}{\partial \theta_i} = -2\Re\left(\frac{\partial \bra{\phi(\vec{\theta})}}{\partial \theta_i}\M\ket{v_0}\bra{v_0}\M^\dag\ket{\phi(\vec{\theta})}\right).
\label{Equ::energyequation}
\end{equation}
The term $\bra{v_0}\M^\dag\ket{\phi(\vec{\theta})}$ can be efficiently measured with the aforementioned method. To measure $\frac{\partial \bra{\phi(\vec{\theta})}}{\partial \theta_i}\M\ket{v_0}$, we have to first decompose the derivative of the state as 
\begin{equation}
    \frac{\partial \ket{\phi(\vec{\theta})}}{\partial \theta_i} = \sum_s f_i^s \ket{\phi_{i}^s(\vec\theta)}.
\end{equation}
Here $\ket{\phi_{i}^s(\vec\theta)} = U_L(\theta_L)\dots \sigma_i^s U_i(\theta_i)\dots U_2(\theta_2)U_1(\theta_1)\ket{0}$, as the derivative of the unitary is decomposed as $\partial U_i(\theta_i)/\partial \theta_i = \sum_sf_i^s\sigma_i^sU_i(\theta_i)$ with tensor products of local operators $\sigma_i^s$ and coefficients $f_i^s$. Then we have $\frac{\partial \bra{\phi(\vec{\theta})}}{\partial \theta_i}\M\ket{v_0}=\sum_s f_i^{s*} \bra{\phi_{i}^s(\vec\theta)}\M\ket{v_0}$ where each term can be efficiently measured with a quantum circuit.
We refer to Appendix~\ref{appendix:circuit}
for the implementation of $\nabla E_{{\M}^{-1}}$. 

\section{Solution verification}
\label{sec:verify}
After finding the solution, it is also practically important to verify whether it is the correct one. This is in general impossible for conventional VQE as neither the ground state nor the ground state energy is known. However, we can verify the solution of the linear algebra problems as the exact solution should always have zero energy. We first consider the matrix-vector multiplication problem. With a solution $\ket{\phi(\vec\theta_{\min})}$, we can estimate $|\bra{\phi(\vec\theta_{\min})}{v_{\M}}\rangle|^2$ via
\begin{equation}
	|\bra{\phi(\vec\theta_{\min})}{v_{\M}}\rangle|^2 = 1 - E_{\M}(\vec\theta_{\min}).
\end{equation}
Therefore, we can also have the fidelity of $\ket{\phi(\vec\theta_{\min})}$ to the exact solution $\ket{v}$  from the energy $E_{\M}(\vec\theta_{\min})$ and the solution can be verified as correct whenever $|\bra{\phi(\vec\theta_{\min})}{v}\rangle|^2$ is close to 1.

For the linear equation, the fidelity $|\bra{\phi(\vec\theta_{\min})}{v_{\M^{-1}}}\rangle|^2$ cannot be determined 
as  $|\bra{\phi(\vec\theta_{\min})}\M^{-1}\ket{v_0}|^2$ or $\bra{v_0}({M}^{-1})^{\dagger}{M}^{-1} \ket{v_0}$ cannot be    measured directly.
Nevertheless, we show that the fidelity is lower bounded by 
\begin{equation}
\label{eq:fidelity_bound}
    |\bra{\phi(\vec\theta_{\min})}{v_{\M^{-1}}}\rangle|^2 \geq 1-\kappa^2 E_{\M^{-1}}(\vec\theta_{\min})
\end{equation}
 with respect to the condition number $\kappa$ and the energy of the Hamiltonian. A similar argument was firstly proposed in Ref.~\cite{bravoprieto2020variational}. The quantum state $\phi(\vec\theta_{\min})$ is close to the true solution when the energy is close to $0$, which provides a criteria for the solution verification. We refer to Appendix \ref{appendix:proof} for details.
An alternative method to verify the solution is by checking whether $\ket{v_0} \propto \M\ket{\phi(\vec\theta_{\min})}$ is satisfied. We can measure ${|\bra{v_0}\M\ket{\phi(\vec\theta_{\min})}|^2}/{\bra{\phi(\vec\theta_{\min})}{\M}^\dag {\M} \ket{\phi(\vec\theta_{\min})}}$ and the  solution is verified as correct when the value is close to $1$. This can be an alternative cost function for VQE.

\section{Optimisation via Hamiltonian morphing}

The VQE method tries to find the global minimum of a large multi-parameter space. For each trial, one starts from a randomly initialised parameter set and obtains a local minimum after several optimisation steps. One can then verify whether the solution is correct by measuring the energy and restart from another random parameter set until the energy is close to 0.  This may need to be repeated many times until a satisfactory solution can be found.  There are several potential issues that can affect the practical performance. First, as the parameter space is large, it may require a large number of repetitions starting from different initial random parameters. This is a common issue in optimisation and machine learning, where a combination of different algorithms may help to speed up the search time. A more serious problem is that the gradient may vanish for randomly initialised parameters in some random quantum circuits of large systems~\cite{mcclean2018barren}. In quantum computational chemistry, physically motivated ans\"atz and a good guess of initial parameters are considered to avoid vanishing gradient, see Refs.~\cite{ourReview,cao2019quantum} for recent reviews. However, this is not straightforward for the linear algebra tasks. 

Here we propose a Hamiltonian morphing optimisation method to avoid these problems. Similar methods have been introduced in Refs.~\cite{PhysRevA.92.042303,saez2018adiabatic} for variational quantum simulation. The morphing method is an analog to adiabatic state preparation, where one slowly varies the Hamiltonian from $H_i$ to $H_f$, as to gradually evolve the corresponding ground state from $\ket{\psi_i}$ to $\ket{\psi_f}$. For our morphing method, we consider the linear equation problem as an example and it works similarly for the matrix-vector multiplication problem. Denote a time-dependent Hamiltonian $H_{\M^{-1}}(t)=\M(t)^{\dagger}(I-\ket{v_0}\bra{v_0})\M(t)$ with $\M(t)$ satisfying $\M(t=0)=I$ and $\M(t=T)=\M$ respectively. Our algorithm starts from the ground state $\ket{v_0}$ of $H_{\M^{-1}}(0)$ and reaches the ground state of the target Hamiltonian $H_{\M^{-1}}(T)$ as follows. Consider discretised time step $\delta t$ and $\M(t)=(1-t/T)I+  t/T\cdot \M$.

\noindent 1. At the $n=0$ step with time $t=0$, we denote the parameters that represent $\ket{v_0}$ as $\vec \theta(0)$, i.e., $\ket{v_0} = \ket{\phi(\vec \theta(0))}$. 

\noindent 2. At the $n^{\textrm{th}}\in [1, T/\delta t)$ step, we use parameters $\vec \theta((n-1)\delta t)$ found in the last step for $H_{\M^{-1}}((n-1)\delta t)$ as the initial position to find parameters $\vec \theta(n\delta t)$ that correspond to the ground state of $H_{\M^{-1}}(n\delta t)$.  

\begin{figure}
\begin{centering}
\includegraphics[width=1\columnwidth]{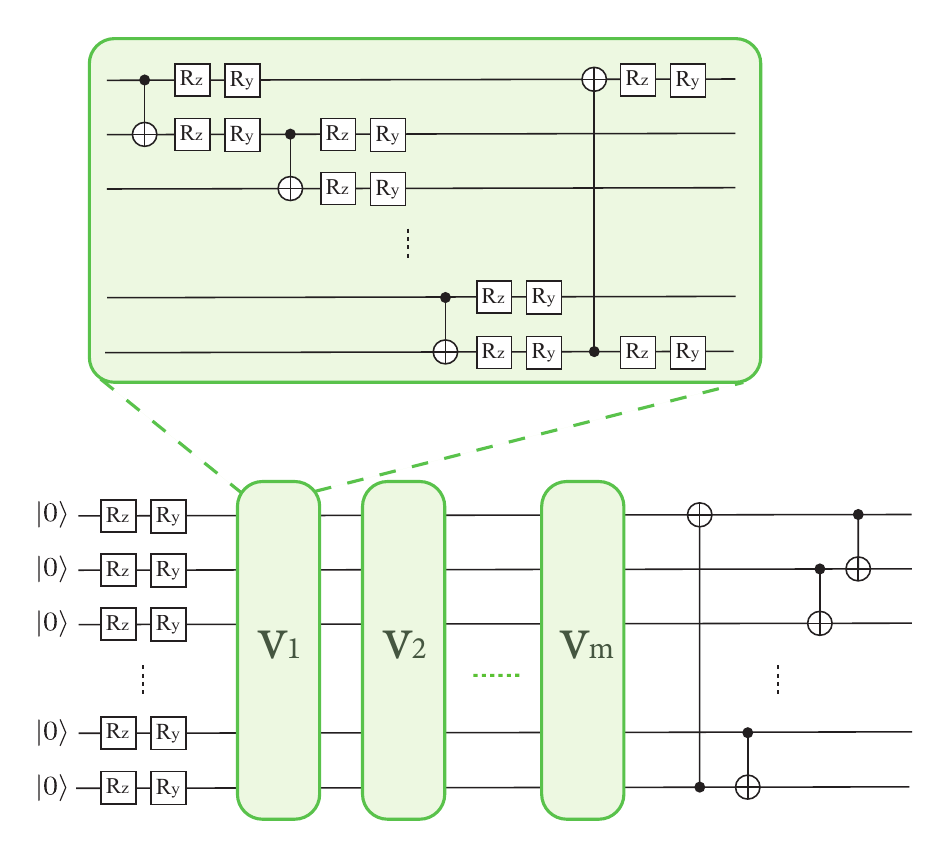}\\
\caption {The circuit ans\"atz. In total, $n$ qubits are required to solve a problem with a  $2^n*2^n$ matrix. Here, $R_y$ and $R_z$ represent single qubit rotations around the $Y$ and $Z$ axes, respectively. The rotation angle (parameter) for each single-qubit gate is initialised from a small random value, which is updated in each of the variational cycles. At the beginning two single-qubit gates are applied to each of the data qubit, followed by sets of CNOT gates, as depicted by the green block $V$. Within each set, a CNOT gate is followed by two single-qubit gates. The number of sets represents the depth of the circuit, which is gradually increased until the desired target state is found. In the end, a set of CNOT gates is applied in reverse order as in $V$. The aim of this set is to ensure that the eigenstate can be found at $t=0$ when the Hamiltonian morphing technique is applied.}
\label{fig:/ansatz}
\end{centering}
\end{figure}

For any positive matrix $M$, when we use a sufficiently small $\delta t$ and a powerful enough ans\"atz, our method is guaranteed to find the exact solution according to the adiabatic theorem~\cite{RevModPhys.90.015002} and it has exponentially speed-up over  classical ones~\cite{PhysRevLett.103.150502}. For a general $\M$, one can introduce an extra qubit so that the energy gap $H_{\M^{-1}}(t)$ is non-vanishing. We refer to Ref.~\cite{PhysRevLett.122.060504} for details. In practice, the initially specified ans\"atz may not be powerful enough to represent the states at all time $t$. Fortunately, when the ans\"atz is insufficient at any time $t$, one can detect it by measuring a high averaged energy. Then one can either simply choose a more powerful ans\"atz or adaptively vary the ans\"atz by changing its structure and adding more gates~\cite{jones2018quantum}. 

\section{Numerical simulation}

We numerically test the variational algorithm for solving linear systems of equations. The simulation is based on the Quantum Exact Simulation Toolkit (QuEST) package~\cite{jones2019quest}, which is a high performance classical simulator of general quantum circuits. 
In our simulation, we consider $2^n$ by $2^n$ positive complex Hermitian matrices $M$ with $n$ qubits. 
Note that even though the matrix size is assumed to be an exponential of 2, an $m$ by $m$ matrix with  $2^{n-1}<m\le2^n$ can be also handled with an $n$-qubit circuit. 
As this work aims to verify the functioning of the algorithm, we consider a general linear algebra task where the matrix is randomly generated with a given condition number $\kappa$ and the input state is also randomly generated.
Here, we use no prior information of the matrix, and thus we consider the ans\"atz as shown in Fig.~\ref{fig:/ansatz}, where each green blocks $V_i$ has a fixed structure. We also vary the number of blocks, called the circuit depth $m$, to endow the ans\"atz with different powers.

We also implement the optimisation based on Hamiltonian morphing with an adaptive ans\"atz. {We divide the total time $T$ into 10 intervals with an equal duration, where $T$ ranges from 20 to 100 depending on the matrix size.} At $t=0$, we set $H_{M^{-1}}=I-|v_0\rangle\langle v_0|$ with the ground state being the input state  $|v_0\rangle$ and start with all single-qubit gates parameterised with a small random rotation angle. In the $i^{\textrm{th}}$ variational cycle, the gradient descent method is used to search for the ground state energy of a time-dependent Hamiltonian $H_{\M^{-1}}(t)=\M(t)(I-\ket{v_0}\bra{v_0})\M(t)$ with Hermitian matrix $\M(t)=(1-i/10)I+i/10\cdot M$, {and a step size $\delta t=0.1$.}
We choose initial parameters to be the ones obtained from the previous cycle.  A small circuit depth $m$ is tried at the beginning and is increased gradually until the fidelity of the qubit state to the target state is higher than 99\%, which is regarded as a success. 

\begin{figure}
\begin{centering}
\includegraphics[width=1\columnwidth]{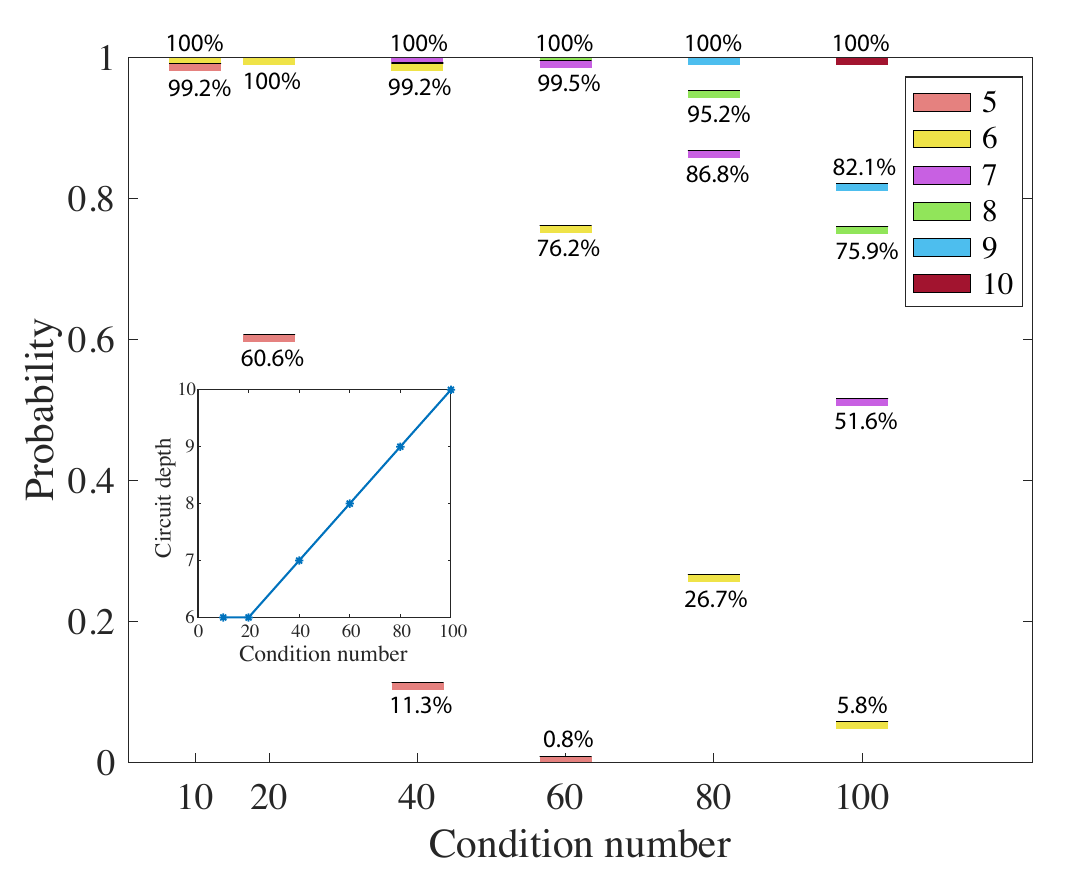}\\
\caption {The probability to find the target state with increasing condition number and circuit depths. This experiment is based on a 6-qubit circuit, thus the matrix size is fixed to be 64
*64. For a given condition number, the success (having a solution with a fidelity higher than 99\%) probability is shown for the circuit with increasing depths, until it reaches 100\% (i.e., all the 4940 runs). The minimum depth required to find the target state with 100\% success probability is plotted in the inserted graph.}
\label{fig:/prob1}
\end{centering}
\end{figure}

\begin{figure}
\begin{centering}
\includegraphics[width=1.04\columnwidth]{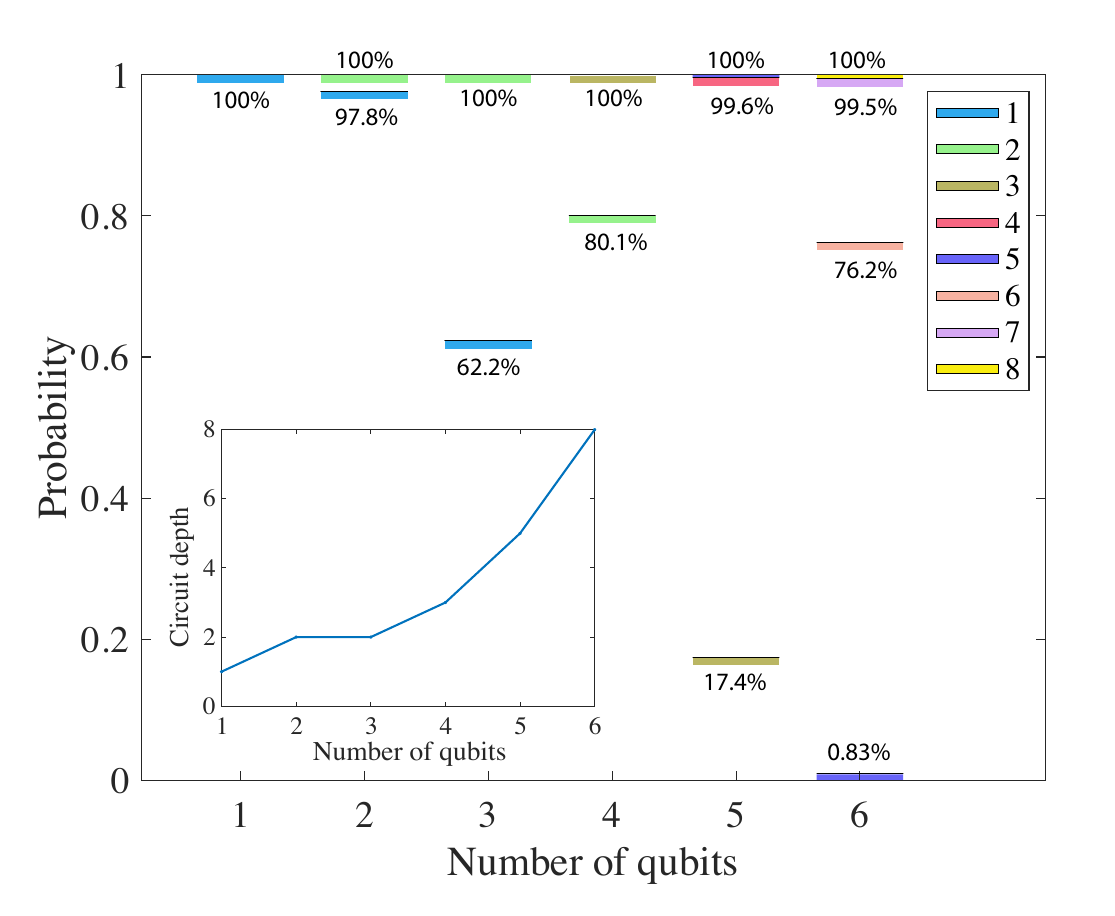}\\
\caption {The probability to find the target state with an increasing number of qubits and circuit depths. For a $n$-qubit system, the condition number is $\kappa=10n$, and the circuit depth is increased gradually until the probability reaches 1. The inserted graph shows the minimum depth found in the simulation when the target state is found with 100\% probability.}
\label{fig:/prob2}
\end{centering}
\end{figure}

We study the complexity of our algorithm with respect to the matrix size and the condition number of the matrix. In matrix inversion problems, the condition number is an important measure that quantifies the sensibility of the solution to perturbations of the input data. The solution is less stable with a larger condition number. 
In this experiment, we define success when the fidelity of the final state is over 99\%.
From Eq.~\eqref{eq:fidelity_bound}, we need to achieve a small energy $E_{M^{-1}}$ to guarantee a high state fidelity for a large condition number $\kappa$.
We first fix the matrix size to be $2^6*2^6$ with $6$ qubits and consider random matrices with different condition numbers.  For different condition numbers from 10 to 100, we  test our algorithm with different circuit depth. For each case, we run 4940 random trials to calculate the averaged success probability with results shown in Fig.~\ref{fig:/prob1}. 
Not surprisingly,  we found that a larger circuit depth ans\"atz generally leads to a higher success probability.
We also obtain the minimum depth required to guarantee the success of finding $|v_{M^{-1}}\rangle$, shown in the figure inset. We find that the minimal depth varies linearly with the condition number in the experiments. In practice, we can design the circuit ans\"atz to increase the representation capability. 

\begin{figure}
\begin{centering}
\includegraphics[width=0.85\columnwidth]{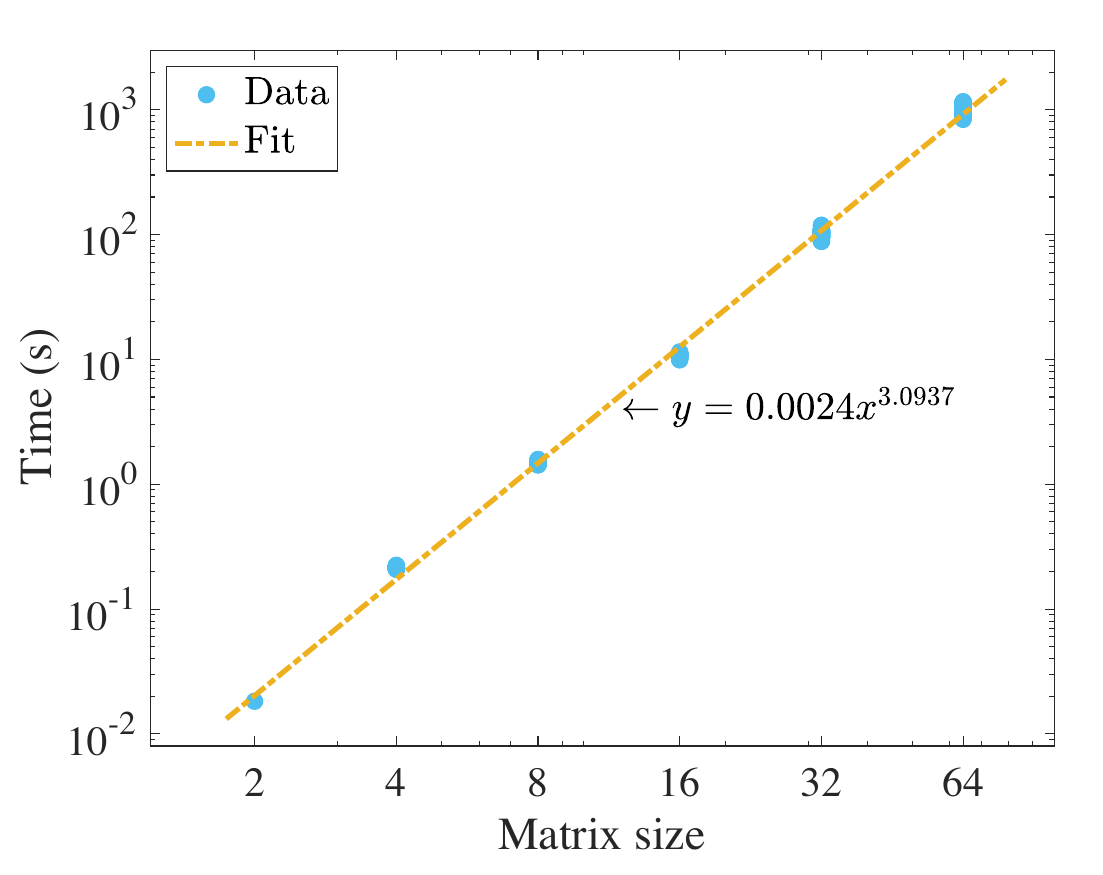}\\
\caption {Time consumed to find the solution with increasing matrix size. For a matrix with size $2^n*2^n$, the condition number is $\kappa=10n$. For a given matrix size, the circuit depth applied is such that the solution is found with 100\% probability (i.e. shown in the inserted graph in Fig.~\ref{fig:/prob2}). The blue dots show the computation time to find the target state. A linear fitted curve (dashed yellow) is also plotted for reference.}
\label{fig:/time}
\end{centering}
\end{figure}

Next, we vary the size $2^n*2^n$ of the matrix $M$, by changing the number of data qubits $n$ in the circuit ans\"atz. For an $n$-qubit system, the condition number is adopted as $\kappa=10n$. 
As shown in Fig.~\ref{fig:/prob2}, the circuit depth leading to a non-zero success probability also increases with respect to the number of qubits. The inserted plot shows the minimum circuit depth for achieving a 100\% success probability. 
Overall, we find a super-linear increase of the circuit depth with respect to the number of qubits. Applied with the minimum circuit depth for 100\% success probability, we now plot the computation time to find the solution in Fig.~\ref{fig:/time}. The log-log scale data points show a roughly linear relation, which is fitted as the yellow dashed curve with the relation $y=0.0024x^{3.1}$, where $y$ is the computation time, and $x$ is the matrix size.
This is not surprising as we consider random matrices with random input states. Even though we consider fixed condition numbers, it only determines properties of eigenvalues of the matrix. For any matrix $M$, all other matrices $UMU^\dag$ with unitary $U$ have the same eigenvalues and hence the same condition number. As the unitary $U$ is an arbitrary unitary, it can lead to a general solution state, which may not be able to be prepared with shallow circuits.  This explains the super-linear dependence of the circuit depth with respect to the number of qubits.

In a more practical scenario, we can assume that we have a sparse decomposition of the matrix or sparse matrices as in Ref.~\cite{PhysRevLett.103.150502}, and the input state is simple to prepare or it is obtained from a previous calculation.
Meanwhile, a better ans\"atz whose design is informed by the matrix can be designed. For example, we can try different ans\"atze, such as the Hamiltonian ans\"atz, defined based on the Pauli decomposition of the matrix.
We also note that the chosen ans\"atz is not optimal for the problem, and a much more compact one can be obtained from circuit compiling~\cite{jones2018quantum} as shown in the Appendix~\ref{appendix:compact}.  As our algorithm has the capability of verifying the solution, it also enables us to dynamically varying the ans\"atz until the solution is found. In our simulation, we dynamically increase the circuit depth and we expect more profound circuit morphing techniques could be designed and exploited.  As this work only aims to verify the basic function and effectiveness of the algorithm for general linear algebra tasks, we leave the simulation of the restricted and practical cases to a future work.

\begin{figure}
\begin{centering}
\includegraphics[width=1\columnwidth]{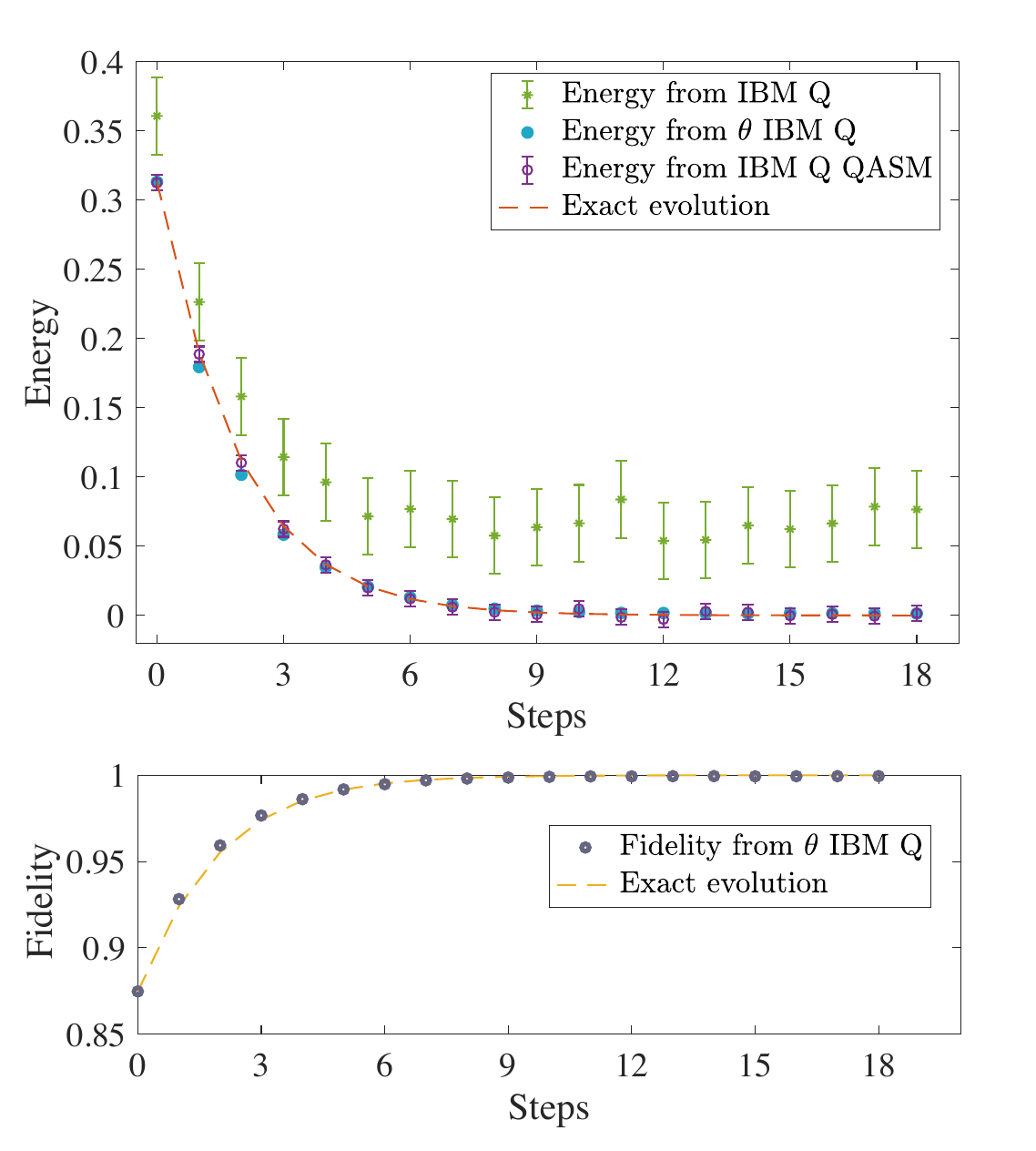}\\
\caption {Energy (upper panel) and state fidelity (lower panel) change with increasing steps. In the upper panel, the green curve shows the measured energy from the IBM quantum processor, while if we put the rotation angle found in each step into a perfect circuit to prepare the trial state, the energy change is depicted by the blue data points. The purple curve refers to the simulation result with perfect circuit implemented in Qiskit. For comparison, the exact energy evolution is plotted as the red dashed curve. In the lower panel, the fidelity of the trial state prepared with the rotation angle discovered (the same set of parameters used to plot the blue curve in the upper panel) is shown as the grey curve. The exact evolution is also shown as a reference.}
\label{fig:/IBMresult}
\end{centering}
\end{figure}

\section{Implementation on the IBM quantum machine}

We implement our variational algorithm using the IBM Q Experience. The processor employed to conduct the experiment is `ibmq-ourense', which has five qubits with $T_2$ time ranging from $20\sim100\mu s$, single-qubit gate error $3.5(\pm 0.5)\times 10^{-4}$, CNOT gate error $8.2 \times 10^{-3}$ and read-out error $2.2\times 10^{-2}$. The circuit is implemented through Qiskit~\cite{aleksandrowicz2019qiskit}, a python-based software development kit for working with OpenQASM and the IBM Q  processors. 

We consider a simple example of solving linear equations with the variational algorithm. The matrix considered is
$M=\left[
  \begin{matrix}
   1.5 & -0.5 \\
   0.5 & 1.5
  \end{matrix} 
  \right]=1.5I-0.5iY,
$ the initial state is $\ket{v_0}=\ket{0}$ and we aim to find $\ket{v_{M^{-1}}}=M^{-1}\ket{v_0}$. The trial state $v_m$ can be prepared with a single-qubit rotation applied on a data qubit. Starting from a small number around zero (0.08 as the case we present here), the rotation angle is updated through $\theta_{i+1} = \theta_{i} - \nabla E_{M^{-1}}(\theta_i)$, where the circuits used to determine $\nabla E_{M^{-1}}$ can be found in Appendix \ref{appendix:circuit}. To keep track of the energy change, the energy in each step is obtained directly by measuring $\langle v_m(\theta_i)|E_{M^{-1}}|v_m(\theta_i)\rangle$ (method shown in Appendix~\ref{appendix:IBM}). The expectation value of each observable is the aggregated result of 8192 runs (the maximum number allowed).

The experimental result is shown in Fig.~\ref{fig:/IBMresult}. Firstly we plot the exact energy evolution as a reference. In the upper panel, as the red dashed curve indicates, the energy decays and gradually approaches 0. The purple data points, which align well with the exact evolution, represent the result simulated in Qiskit with perfect circuits each evaluated 200k times. The experimentally measured energy change is plotted as the green curve. We see a fast decay followed with fluctuations between 0.05 and 0.1. The deviation mostly results from noise within the circuit, especially the notable readout error. However, if we calculate the energy directly using the rotation angle discovered in the same experiment, as depicted by the blue data points, the energy change nearly fits the exact evolution. As a comparison, we also apply the the same set of experimentally obtained rotation angles to prepare the state $\ket{v_m}$ and plot the fidelity ($|\langle v_m|v_{M^{-1}}\rangle|^2$) change, as shown by the grey data points in the lower panel of Fig.~\ref{fig:/IBMresult}. We see they nearly overlap with the orange curve and lead to a very high fidelity of 99.95\%. The result suggests that our algorithm works with a small noisy device, and it can be foreseen that solving linear systems with higher dimensions can be achievable with quantum devices with higher gate and measurement fidelities.

\section{Applications}

In this section, we discuss applications of our algorithm, including Hamiltonian simulation and open systems simulation. 

We first show the application for variational Hamiltonian simulation. 
Starting from $\ket{\psi_0}$, the Hamiltonian simulation task is to prepare the state $\ket{\psi_t} = e^{-iHt}\ket{\psi_0}$ at time $t$ with sparse Hamiltonian $H$. For instance, when the Hamiltonian can be decomposed as a linear sum of polynomial terms $H=\sum_j \lambda_j \sigma_j$, we can make use of the Trotterization method to have 
\begin{equation}
	\ket{\psi_t} = \left(\prod_ie^{-i\lambda_j \sigma_jt/K}\right)^K\ket{\psi_0} + \mathcal{O}(t^2/K),
\end{equation}
with $K$ Trotter steps. As the circuit depth increases with the evolution time, the circuit depth can be large for long evolution time. The variational quantum algorithm for simulating real-time dynamics with a constant circuit depth has been proposed by Li and Benjamin~\cite{Li2017}. Here, we present an alternative algorithm based on matrix-vector that can be more robust to the one in Ref.~\cite{Li2017}. The key idea of Ref.~\cite{Li2017} is to prepare the state at any time $t$ via a parameterised state $\ket{\phi(\vec\theta(t))}$ with parameters $\vec \theta = (\theta_1,\theta_2,...,\theta_L)$,  and update the parameters from $\vec\theta(t)$ to $\vec\theta(t+\delta t)$ that effectively realise $e^{-iH\delta t}\ket{\phi(\vec\theta(t))}$ via a deterministic process $\vec\theta(t+\delta t) = \vec\theta(t) + \dot{\vec{\theta}}(t) \delta t$ with $\dot{\vec{\theta}}(t) = A^{-1}C$, where $A_{i,j}(t)=-\textrm{Im}\left(\frac{\partial\bra{\phi(\vec{\theta})}}{\partial\theta_i} \frac{\partial\ket{\phi(\vec{\theta})}}{\partial \theta_j}\right)$ and $C_i(t)=\textrm{Re}\left(\frac{\partial\bra{\phi(\vec{\theta})}}{\partial\theta_i} H \ket{\phi(\vec{\theta})}\right)$. Each term of $A$ and $C$ can be measured via a quantum circuit. 
        
Instead of deterministically obtaining $\vec\theta(t+\delta t)$ via $\dot{\vec{\theta}}(t)$, we make use of the algorithm introduced in this paper. That is, to realise $e^{-iH\delta t}\ket{\phi(\vec\theta(t))}$, we can consider $\M = e^{-iH\delta t}\approx 1 - iH\delta t$ and update the parameters as $\ket{\phi(\vec\theta(t+\delta t))} = \M\ket{\phi(\vec\theta(t))}$. {Therefore, for each time step, we can update the parameters according to the matrix-vector algorithm. 
We can verify the solution as discussed in the section~\ref{sec:verify}.
We note that the number of measurements at each time is reduced from $\mathcal{O}(L^2)$ to $\mathcal{O}(L) $.}
This can be similarly applied for the imaginary time evolution which aims to realise $\ket{\psi_{\tau}} = e^{-H\delta \tau}\ket{\psi_{0}}$.
The matrix vector multiplication algorithm could be leveraged for these dynamics simulation tasks, which are in general challenging for classical computers to simulate real and imaginary time evolution.
 
Our algorithms is efficiently applicable to problems when the initial state vector $v_0$ is a complicated many-qubit state. In this case, we can even consider a  simple sparse matrix $M$, such as a single Pauli matrix. While such a simple $M$ and complicated  $v_0$ might not be a common classical linear algebra task, it could appear in certain quantum computing subroutines. For example, to simulate open quantum system evolution using variational algorithms, we need to apply some jump operators on the quantum state~\cite{endo2020variational}.
More specifically, considering Lindblad master equation $d_t \rho =-i[H, \rho]+\mathcal{L}\rho$, 
it can be equivalently described by the stochastic Schr\"odinger equation, which averages trajectories of pure
state evolution under continuous measurements.
The whole process consists of two two parts: 
the continuous generalised time evolution  and the quantum jump process.
The simulation of the quantum jump process requires the matrix-vector multiplication, which in general could be hard for classical simulation. Our algorithms can be leveraged to simulate this process. We refer to Appendix for more details.

While in this work, we focus on the two problems of matrix-vector multiplication and matrix inversion, other popular linear algebra problems, like singular value decomposition, could also be solved by the variational algorithm with a proper cost function, as demonstrated by some recent works~\cite{bravo2020quantum, wang2020variational}.






\vspace{0.2cm}

\section{Discussion}
In this work we present variational algorithms for solving linear algebra problems including linear system of equations and matrix-vector multiplications with NISQ devices. 
We design an effective Hamiltonian with the ground state as the solution of the target problem, and solve it with the recently-proposed variational algorithms. As the ground state has zero energy, it also enables us to verify the solution and adjust the ans\"atz dynamically until the desired accuracy is achieved.  We also introduce the Hamiltonian morphing technique to avoid local minima and to accelerate the search progress, and numerically verify our algorithm for solving the linear systems of equations. Owing to computation limitations, we consider randomly chosen matrices with a small range of matrix size. Within this range and with the non-optimal ans\"atz, the result suggests that the circuit depth based on our algorithm scales linearly with the condition number and super-linearly with the number of qubits.
The numerical simulations verify the basic function of the algorithm for matrix inversion tasks with general matrices and input states.
We note that it can be challenging for solving linear systems of equations in general, and it is an interesting direction to discuss the potential speed-up for the restricted and practical tasks. 
A proof-of-principle experiment conducted on the IBM Q device is presented. In the presence of significant noise including very appreciable readout error, the optimised parameter corresponds to a state with 99.95\% fidelity, which further verified our algorithms are compatible with NISQ hardware. 
We expect further improvements of the performance of the algorithm for sparse matrices together with more compact and sophisticated ans\"atze.  We also expect that our results may shed light on quantum machine learning and quantum optimisations in the NISQ era. \\

\noindent\emph{Note added.}---
Other related works have appeared around the same time since  our paper was posted on arxiv~\cite{huang2019near,an2019quantum,bravoprieto2020variational}.
These works discussed variational quantum algorithms for solving linear systems of equations, and our results can be compared.

\section*{Acknowledgements}
This work is supported by the EPSRC National Quantum Technology Hub in Networked Quantum Information Technology (EP/M013243/1). SE is supported by Japan Student Services Organization (JASSO) Student Exchange Support Program (Graduate Scholarship for Degree Seeking Students). YL is supported by NSAF (Grant No. U1730449). SCB and XX acknowledge support from the European Quantum Technology Flagship project AQTION. We acknowledge use of the IBM Q for this work. The views expressed are those of the authors and do not reflect the official policy or position of IBM or the IBM Q team.

\vspace{0.3cm}
\section*{AUTHOR CONTRIBUTIONS}
Xiao Yuan, Xiaosi Xu and Jinzhao Sun conceived the project. Xiaosi Xu and Jinzhao Sun carried out the numerical simulation.  Xiaosi Xu and Xiao Yuan performed the experiment on IBM Q. Simon C.~Benjamin and Xiao Yuan supervised the project. All authors discussed the results and contributed to the writing of the paper.

\section*{DATA AVAILABILITY}
The data sets generated to support the findings of this study are available from the corresponding author upon reasonable request.

\section*{COMPETING INTERESTS}
The authors declare no competing interests.

\bibliographystyle{SciAdv}
\bibliography{bib.bib}

\clearpage
\widetext

\appendix

\section{The lowest eigenstate of $H_{{\M}}$ and $H_{{\M}^{-1}}$ and solution verification}
\label{appendix:proof}
We first show the lowest eigenstate of $H_{{\M}}$ is $\ket{v_M} = M \ket{v_0}$. 
For an arbitrary vector $\ket{\varphi}$, we have
\begin{align}
\bra{\varphi}H_{\M}\ket{\varphi}
&= \bra{\varphi}(I-\frac{{\M}\ket{v_0}\bra{v_0}{\M^\dag}}{\|{\M} \ket{v_0}\|^2}) \ket{\varphi} \\
&=1- \frac{ |\bra\varphi{\M}\ket{v_0}|^2 }{\|{\M} \ket{v_0}\|^2} \geq 0
\end{align}
It is easy to check that $H_{\M} \ket{v_M} =0$, and thus $\ket{v_M}$ is the lowest eigenvector of $H_{{\M}^{-1}}$. 

Similarly, we show the lowest eigenstate of $H_{{\M}^{-1}}$ is $\ket{\psi_0} =  {\M}^{-1} \ket{v_0}$. For an arbitrary vector $\ket{\varphi}$, 
\begin{align}
\bra{\varphi}H_{{\M}^{-1}}\ket{\varphi}
&= \bra{\varphi}{\M} (I-\ket{v_0}\bra{v_0}){\M}\ket{\varphi} \\
&=\sum_i |\bra{\varphi}{\M}\ket{\varphi_i} |^2 \geq 0, 
\end{align}
where we denote $I-\ket{v_0}\bra{v_0}=\sum_i \ket{\varphi_i}\bra{\varphi_i}$, $\braket{\varphi_i|\varphi_j}=\delta_{i,j}$. Therefore $H_{{\M}^{-1}}$ is a positive semidefinite matrix. On the other hand, $H_{{\M}^{-1}} {\M}^{-1} \ket{v_0} =0$, thus ${\M}^{-1} \ket{v_0}$ is the lowest eigenvector of $H_{{\M}^{-1}}$. As such, the linear algebra task is converted to the task of finding the ground state of the Hamiltonian, which can be accomplished by various methods, 
such as variational quantum eigensolver, adaptive variational algorithms, global search algorithms, imaginary time evolution algorithms, etc. We mainly focus on the variational quantum eigensolver in the main text, and we discuss  the approach of imaginary time evolution approach in the next section. 

In general we cannot verify the solution of general many-body Hamiltonian owing to unknown ground state and energy, but  we can  verify the solutions of the two linear algebra tasks considered in the main text.  In the following, we show more details for the derivation of Eq.~(14) in the main text.
The variational state obtained from the parameterised circuits can be decomposed into the eigenstate basis of the Hamiltonian as 
$\ket{\phi} = \sum_j c_j \ket{\psi_j}$, where we denote  energy eigenstates $\ket{\psi_j}$ and the eigenenergies $E_j$ ($E_0 = 0$), and  we assume the eigenstates are non degenerate for simplicity. 
The state fidelity is $\braket{\phi|v_{\M^{-1}}} = |c_0|^2$.
The energy has the lower bound as 
\begin{equation}
    E_{M^{-1}} = \sum_{j \geq 1} |c_j|^2 E_j \geq \sum_{j \geq 1} |c_j|^2 E_1.
\end{equation}
The bound saturates when there are only spectral weights on  ground state and first excited state.
From Ref. \cite{PhysRevLett.122.060504}, the energy gap satisfies $E_1 \geq 1/\kappa^2$. Therefore, with the normalisation $\sum_j |c_j|^2 = 1$, we have the lower bound of solution fidelity as
\begin{equation}
    |\bra{\phi(\vec\theta_{\min})}{v_{\M^{-1}}}\rangle|^2 \geq 1-\kappa^2 E_{\M^{-1}} 
\end{equation} with respect to the the condition number and the energy 
of the Hamiltonian.
Similar discussions can be found in a recent posted paper \cite{bravoprieto2020variational}.

\section{Optimisation methods searching for $\vec\theta_{min}$}
\label{appendix:ITE}
Search for the optimal set of parameters based on the cost function can be generalised as an optimisation problem, which can be handled with many methods. We describe the approach of imaginary time evolution approach~\cite{endo2020variational}, which is equivalent to the quantum natural gradient descent for the pure state case. Based on the McLachlan' s variational principle, the minimisation starts with a initial guess of parameters $\vec\theta_0$, which is to be updated in each cycle with its derivative given by:
\begin{equation}
    \sum_j M_{i,j}\dot{\theta_j}=-V_i,
\end{equation}
where 
\begin{equation}
\begin{aligned}    &M_{i,j}=\Re\left(\frac{\partial \langle\phi(\vec\theta)|}{\partial\theta_i}\frac{\partial |\phi(\vec\theta)\rangle}{\partial\theta_j}\right),
    &V_i=\Re\left(\frac{\partial \langle\phi(\vec\theta)|}{\partial\theta_i}H|\phi(\vec\theta)\rangle\right).
\end{aligned}
\end{equation}
The Hamiltonian $H$ for the two cases described in the main text would be $H_M$ and $H_{M^{-1}}$ respectively. 
We note that a quantum state $\ket{\psi}$ evolves in imaginary time $\tau$ can be expressed as $\ket{\psi(\tau) }= \sum_j c_j e^{-E_j}\tau \ket{e_j}$,
where the spectral weight of energy eigenstates $\ket{e_j}$ is suppressed exponentially by their eigenenergies $E_j$.
Therefore,  along the imaginary time evolution trajectory,  the lowest eigenstate of Hamiltonian $H$ can be determined in the long time limit.
Therefore, by updating $\vec\theta$ via:
\begin{equation}
\vec\theta_{i+1}=\vec\theta_i+\Delta \tau \dot{\vec{\theta_i}}
\end{equation}
with a small time step $\Delta \tau $, we can obtain the ground state of the Hamiltonian, i.e., the solutions ${\M} \ket{v_0}$ or ${\M}^{-1} \ket{v_0}$.


\section{Measuring general matrices and resource estimation}
\label{appendix:sparse}
Here, we show how to measure a general matrix
\begin{equation}\label{Eq:appexpand}
    M = \sum_j \lambda_j \sigma_j,
\end{equation}
with the assumption that we could efficiently sample according to the distribution $p_j= |\lambda_j|/\sum_j|\lambda_j|$. 
It includes the case where the summation has polynomial number of terms discussed in the main text, as well as the case where the summation has exponential number of terms. For example, when $M$ is a sparse matrix, i.e., there are only a polynomial number of non-zeros terms of the matrix elements, 
\begin{equation}
    \{({\bf x}^{(i)}, {\bf y}^{(i)}, M_{{\bf x}^{(i)},{\bf y}^{(i)}})\},
\end{equation}
where the $(i)$ represents the index of non-zero terms and $M_{{\bf x}^{(i)},{\bf y}^{(i)}}$ represents the matrix element at $({\bf x}^{(i)},{\bf y}^{(i)})$. Suppose $M$ is an $N\times N$ matrix with $n = \lceil\log_2 N\rceil$ and a binary representation of ${\bf x}^{(i)}$ and ${\bf y}^{(i)}$ as
\begin{equation}
\begin{aligned}
    {\bf x}^{(i)} &= (x_1^{(i)}, x_2^{(i)}, \dots, x_n^{(i)}), \\
    {\bf y}^{(i)} &= (y_1^{(i)}, y_2^{(i)}, \dots, y_n^{(i)}).
\end{aligned}
\end{equation}
Now we can represent $M$ as
\begin{equation}
    \begin{aligned}
    M &= \sum_{(i)} M_{{\bf x}^{(i)},{\bf y}^{(i)}} \ket{{\bf x}^{(i)}}\bra{{\bf y}^{(i)}},\\
    &= \sum_{(i)} M_{{\bf x}^{(i)},{\bf y}^{(i)}} \ket{ x_1^{(i)}}\bra{y_1^{(i)}}\otimes \ket{x_2^{(i)}}\bra{ y_2^{(i)}}\otimes \dots\otimes \ket{ x_n^{(i)}}\bra{y_n^{(i)}}.
    \end{aligned}
\end{equation}
Note that each $\ket{ x_j^{(i)}}\bra{y_j^{(i)}}$ could be expanded as a linear combination of two unitary operators. In particular, we have
\begin{equation}
\begin{aligned}
    \ket{0}\bra{0} &= (I+Z)/2,\\
    \ket{0}\bra{1} &= (X-iY)/2,\\
    \ket{1}\bra{0} &= (X+iY)/2,\\
    \ket{1}\bra{1} &= (I-Z)/2,\\
\end{aligned}
\end{equation}
and the matrix $M$ could be expanded as the form of Eq.~\eqref{Eq:appexpand}. 
Specifically, we denote 
\begin{equation}
    \ket{{\bf x}^{(i)}}\bra{{\bf y}^{(i)}} = \frac{1}{2^n}\sum_{j^{(i)}} \sigma_{j^{(i)}},
\end{equation}
with $\sigma_{j^{(i)}}$ being one of $2^n$ expanded unitary operators. Then we have
\begin{equation}
    M = \sum_{(i)}M_{{\bf x}^{(i)},{\bf y}^{(i)}} \sum_{j^{(i)}} \frac{1}{2^n} \sigma_{j^{(i)}}.
\end{equation}
To randomly measure $M$, we can first sample $(i)$ according to $q(i) =  |M_{{\bf x}^{(i)},{\bf y}^{(i)}}| / \sum_{(i)}  |M_{{\bf x}^{(i)},{\bf y}^{(i)}}| $ and then toss $n$ unbiased coins to get $j^{(i)}$ for each $\sigma_{j^{(i)}}$. The phase of $M_{{\bf x}^{(i)},{\bf y}^{(i)}}$ could be absorbed into $\sigma_{j^{(i)}}$. 

We note that the sample complexity is related to $C = \sum_j {|\lambda_j|}$. In particular, according to Hoeffding's inequality, we need $\mathcal O(\log(\delta^{-1})C^2/\varepsilon^2)$ samples to measure $\bra{\phi}M \ket{\psi}$ to an accuracy $\varepsilon\in(0,1)$ with failure probability $\delta\in(0,1)$. The cost is small when our assumption holds. Specifically, $C$ increases polynomially with respect to the system size when $M$ is expanded as a polynomial number of unitary operators. When $M$ is a sparse matrix that has polynomial number of nonzero terms, we have $C = \sum_j {|\lambda_j|} = \sum_{(i)}  |M_{{\bf x}^{(i)},{\bf y}^{(i)}}| $, which also leads to a polynomial resource cost for measurements.

\section{Circuit implementation of $\nabla E_{M^{-1}}$}
\label{appendix:circuit}
Here we show that the circuits to implement $\nabla E_{M^{-1}}$ can be found similarly as we do for $\nabla E_M$. 
We have \begin{align}
H_{{\M}^{-1}}={\M}^\dag (I-\ket{v_0}\bra{v_0}){\M},
\end{align}
thus 
\begin{equation}
\begin{aligned}
\frac{\partial E_{M^{-1}}}{\partial \theta_i}&=\frac{\partial}{\partial \theta_i}(\bra{\phi(\vec\theta)}H_{M^{-1}}\ket{\phi(\vec\theta)})=2\Re\left(\frac{\partial\bra{\phi(\vec\theta)}}{\partial \theta_i}H\ket{\phi(\vec\theta)}\right)\\
&=2\Re\left(\frac{\partial\bra{\phi(\vec\theta)}}{\partial \theta_i}M^\dag M\ket{\phi(\vec\theta)}\right)-2\Re\left(\frac{\partial\bra{\phi(\vec\theta)}}{\partial\theta_i}M^{\dagger}\ket{v_0}\bra{v_0}M\ket{\phi(\vec\theta)}\right).
\end{aligned}
\label{Equ::energyinverse}
\end{equation}

\begin{figure}
\begin{centering}
\includegraphics[width=0.8\columnwidth]{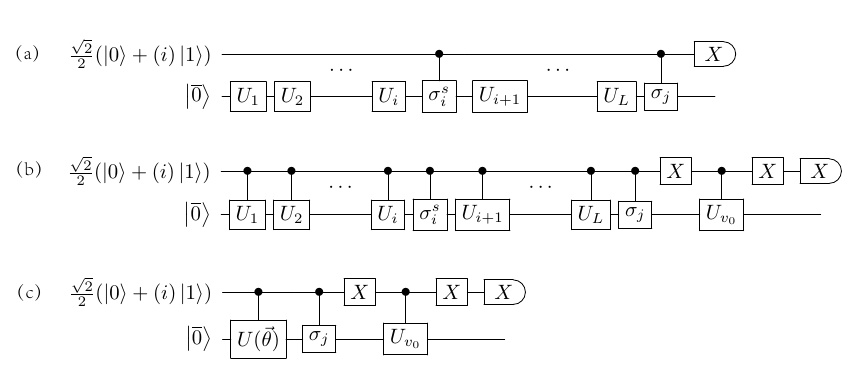}\\
\caption {The circuits to implement $\nabla E_{M^{-1}}$. The first qubit is the ancilla, and the second refers to the data qubit register. The real part can be obtained from measuring $\braket{X}$ when the ancilla initialised at $\frac{\sqrt{2}}{2}(\ket{0}+\ket{1})$, while the imaginary part is evaluated if the ancilla is initialised at $\frac{\sqrt{2}}{2}(\ket{0}+i\ket{1})$.}
\label{fig:circuitimp}
\end{centering}
\end{figure}

Suppose $M$ can be decomposed into combinations of Pauli terms $M=\sum_j\alpha_j$, $M^{\dag}M$ can then be expressed as $M^{\dag}M=\sum_j\beta_j\sigma_j$, where the coefficients $\alpha$ and $\beta$ for each term are different. We decompose the derivative of the state as 
\begin{equation}
    \frac{\partial \ket{\phi(\vec{\theta})}}{\partial \theta_i} = \sum_s f_i^s\ket{\phi_{i}^s(\vec\theta)},
\end{equation}
the same way as described in the main text. Then the first term in Eq.~\eqref{Equ::energyinverse} is expressed as

\begin{equation}
\begin{aligned}
\frac{\partial \bra{\phi(\vec{\theta})}}{\partial \theta_i}M^{\dag}M\ket{\phi(\vec{\theta})}&=\sum_s f_i^{s*} \bra{\phi_{i}^s(\vec\theta)}M^{\dag}M\ket{\phi(\vec\theta)}\\
&=\sum_j\sum_s c_i^{s,j}\bra{0}U_1^{\dag}(\theta_1)U_2^{\dag}(\theta_2)\dots U_i^{\dag}(\theta_i)\sigma_i^s\dots U_L^{\dag}(\theta_L)\sigma_jU_L(\theta_L)\dots U_2(\theta_2)U_1(\theta_1)\ket{0}.
\end{aligned}
\end{equation}
The real and imaginary value of each term $\bra{0}U_1^{\dag}(\theta_1)U_2^{\dag}(\theta_2)\dots U_i^{\dag}(\theta_i)\sigma_i^s\dots U_L^{\dag}(\theta_L)\sigma_jU_L(\theta_L)\dots U_2(\theta_2)U_1(\theta_1)\ket{0}$ can be evaluated with the circuit shown in Figure~\ref{fig:circuitimp}(a), by initialising the ancilla qubit to be at $\frac{\sqrt{2}}{2}(\ket{0}+\ket{1})$ and $\frac{\sqrt{2}}{2}(\ket{0}+i\ket{1})$ respectively.

The second term can be separated as two parts, $\frac{\partial\bra{\phi(\vec\theta)}}{\partial\theta_i}M^{\dagger}\ket{v_0}$ and $\bra{v_0}M\ket{\phi(\vec\theta)}$. The first part can be expressed as 
\begin{equation}
\begin{aligned}
\frac{\partial\bra{\phi(\vec\theta)}}{\partial\theta_i}M^{\dagger}\ket{v_0}&=\sum_s f_i^{s*}\bra{\phi_{i}^s(\vec\theta)}M^{\dag}U_{v_0}\ket{0}\\
&=\sum_j\sum_s d_i^{s,j}\bra{0}U_1^{\dag}(\theta_1)U_2^{\dag}(\theta_2)\dots U_i^{\dag}(\theta_i)\sigma_i^s\dots U_L^{\dag}(\theta_L)\sigma_jU_{v_0}\ket{0},
\end{aligned}
\end{equation}
where each term can be evaluated with the circuit shown in Figure~\ref{fig:circuitimp}(b).
The second part is written as 
\begin{equation}
\bra{v_0}M\ket{\phi(\vec\theta)}=\bra{0}U_{v_0}^\dag M U(\vec\theta)\ket{0}=\sum_jk_j\bra{0}U_{v_0}^\dag\sigma_jU(\vec\theta)\ket{0}
\end{equation}
where each term can be evaluated with the circuit shown in Figure~\ref{fig:circuitimp}(c).

\section{Example with a more compact circuit}
\label{appendix:compact}
In this paper, we apply the hardware-efficient ans\"atz for practical soundness, however, this is not the optimal ans\"atz for most cases. Here we take one case as an example, where the solution is found difficult to reach in our numerical simulations. 

\begin{figure}
\begin{centering}
\includegraphics[width=1\columnwidth]{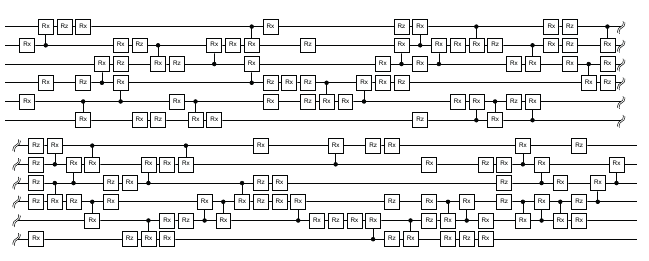}
\caption {A more compact circuit found with quantum compilation to realise the target state. Each single- and multi-qubit gate is controlled with one parameter. In total 126 parameters are included in the circuit, which realises the state with a fidelity higher than 99.9\%.  }
\label{fig:circuitrec}
\end{centering}
\end{figure}

The matrix in this case is a complex matrix with size of 64*64, and condition number $\kappa=60$. $\ket{v_0}$ is a complex vector, and the target state is $\ket{v_{\M^{-1}}} = \frac{\M^{-1} \ket{v_0}}{\|{\M}^{-1} \ket{v_0} \|}$. It is found in the numerical simulation that the target state can not be represented by a circuit with depth 7 but can by a circuit with depth 8. 
On the other hand, applying quantum compiling, it is found that the target state can be realised with a circuit that requires only half the number of parameters. Figure~\ref{fig:circuitrec} is one example of the circuits found to prepare a state with fidelity higher than 99.9\% compared with the target state. A limited topology with only nearest-neighbor interactions is considered here as it is more hardware efficient. The detailed description and the parameters can be found at https://questlink.qtechtheory.org/.

\section{Circuits implemented on the IBM Quantum processor}
\label{appendix:IBM}
\begin{figure}
\begin{centering}
\includegraphics[width=0.7\columnwidth]{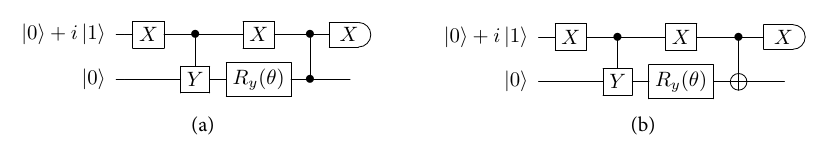}\\
\caption {Circuits to evaluate (a) $\Im(\bra{0}Ye^{-i\frac{\theta}{2}Y}Ze^{i\frac{\theta}{2}Y}Y\ket{0})$ and (b) $\Im(\bra{0}Ye^{-i\frac{\theta}{2}Y}Xe^{i\frac{\theta}{2}Y}Y\ket{0})$ respectively.}
\label{fig:/circuitExp}
\end{centering}
\end{figure}

We consider solving the linear equation $\ket{v_{\M^{-1}}} = \frac{\M^{-1} \ket{v_0}}{\|{\M}^{-1} \ket{v_0} \|}$ with the matrix $M=\left[
  \begin{matrix}
  1.5 & -0.5 \\
  0.5 & 1.5
  \end{matrix} 
  \right]=1.5I-0.5iY,
$ and $\ket{v_0}=\ket{0}$. To simplify the problem, we first write the Hamiltonian into a linear combination of Pauli terms. As $\ket{0}\bra{0}=0.5*(X+Z)$, the Hamiltonian $H_{{\M}^{-1}}={\M}^\dag (I-\ket{v_0}\bra{v_0}){\M}$ can be expressed as $H_{{\M}^{-1}}=(1.5I-0.5iY)*(1.5I+0.5iY)-0.5*(1.5I+0.5iY)(X+Z)(1.5I-0.5iY)=1.25I-Z+0.75X$. We suppose the trial state is prepared by applying a single-qubit $Y$ rotation such as $\ket{\phi(\theta)}=e^{i\frac{\theta}{2}Y}\ket{0}$. Therefore, the derivative of the energy 
\begin{equation}
\begin{aligned}
\frac{\partial E_{M^{-1}}}{\partial \theta}&=\frac{\partial}{\partial \theta}(\bra{\phi(\theta)}H_{M^{-1}}\ket{\phi(\theta)})=2\Re\left(\frac{\partial\bra{\phi(\theta)}}{\partial \theta}H_{M^{-1}}\ket{\phi(\theta)}\right)\\
&=\Im\left(\bra{0}Ye^{-i\frac{\theta}{2}Y}(1.25I-Z+0.75X)e^{i\frac{\theta}{2}Y}Y\ket{0}\right)\\
&=1.25-\Im\left(\bra{0}Ye^{-i\frac{\theta}{2}Y}Ze^{i\frac{\theta}{2}Y}Y\ket{0}\right)+0.75\Im\left(\bra{0}Ye^{-i\frac{\theta}{2}Y}Xe^{i\frac{\theta}{2}Y}Y\ket{0}\right),
\end{aligned}
\label{Equ::energyinverse}
\end{equation}
where the second and the third terms can be evaluated with the circuits shown in Figure.~\ref{fig:/circuitExp}(a) and (b) respectively.

We also measure the energy in each step to keep track of the energy change. As $E=\langle v_m|H|v_m\rangle=\langle v_m|(1.25I-Z+0.75X)|v_m\rangle$, we obtain $E$  by measuring $\langle v_m|Z|v_m\rangle$ and $\langle v_m|X|v_m\rangle$ directly.

\section{Simulation of open system dynamics}

We briefly review the simulation of open quantum system dynamics introduced in Ref.~\cite{endo2020variational}.
We consider to simulate the open system discribed by the Lindblad master equation as 
\begin{align} 
\frac{d}{dt} \rho =-i[H, \rho]+\mathcal{L}[\rho]. 
\label{eq:lindblad}
\end{align}
where $H$ is the system Hamiltonian and $\mathcal{L}[\rho] =\sum_k \frac{1}{2}(2 L_k \rho L_k ^\dag -L_k^\dag L_k \rho -\rho L_k ^\dag L_k)$ describes the interaction with the environment
with Lindblad operators $L_k$. 
Eq.~\eqref{eq:lindblad} can be equivalently described by the stochastic Schr\"odinger equation, which averages trajectories of pure
state evolution under continuous measurements.
Each single trajectory $\ket{\psi(t)}$ can be expressed by  
 \begin{equation}
     d\ket{\psi(t)}= {A} \ket{\psi(t)} dt   +\sum_{k } {B}_k  dN_k.
 \end{equation}
 Here, the whole process consists of two
 two parts: 
the continuous generalised time evolution $A=-iH-\frac{1}{2}\sum_{k}( L_k^\dag L_k -\braket{L_k^\dag L_k }$ and the quantum jump process $B_k= \frac{L_k \ket{\psi(t)}}{||L_k \ket{\psi(t)}||} - \ket{\psi(t)}$ with $dN_k$ randomly takes either $0$ or $1$. 
For the quantum jump process, the state discontinuously jumps to  $L_k \ket{\psi(t)}/ \|L_k \ket{\psi(t)} \|$ with probability $E[dN_k]$. 
We can determine the time that quantum jump happens (See Ref.~\cite{endo2020variational} for details), and realise 
the jump process   by the matrix-vector multiplication proposed in this work. Our method provides an alternative solution for the open system simulation.

\end{document}